\documentclass[10pt,a4paper,notitlepage]{article}

\usepackage{graphicx}% Include figure files
\usepackage{dcolumn}% Align table columns on decimal point
\usepackage{bm}% bold math
\usepackage{subeqnarray}
\date{ }

\begin{document}

\title{\rm \textbf{Spheroidal and elliptical galaxy radial velocity dispersion determined from Cosmological General Relativity}  }
\author{\textbf{John G. Hartnett}\\
School of Physics, the University of Western Australia,\\
 35 Stirling Hwy, Crawley 6009 WA Australia\\
\textit{john@physics.uwa.edu.au}}

\def\kmsmpc{${km ~s^{-1}Mpc^{-1}}$}

\maketitle

\begin{abstract}
Radial velocity dispersion in spheroidal and elliptical galaxies, as a function of radial distance from the center of the galaxy, has been derived from Cosmological Special Relativity. For velocity dispersions in the outer regions of spherical galaxies, the dynamical mass calculated for a galaxy using Carmelian theory may be 10 to 100 times less than that calculated from standard Newtonian physics. This means there is no need to include halo dark matter. The velocity dispersion is found to be approximately constant across the galaxy after falling from an initial high value at the center. 

\end{abstract}


\section{Introduction}
The motion of stars or the motion of gases as characterized by the spectroscopic detection of neutral hydrogen and other gases in spheroidal and elliptical galaxies has caused concern for astronomers for many decades. Newton's law of gravitation predicts much lower radial velocity dispersions than those measured. This has led to the assumption of the existence of halo `dark matter' surrounding galaxies but transparent to all forms of electromagnetic radiation. 

Carmeli \cite{Carmeli2000,Carmeli2002} formulated an extension of Einstein's general theory, in an expanding universe taking into account the Hubble expansion as a fundamental axiom, which imposes an additional constraint on the dynamics of particles \cite{Carmeli1982}. Carmeli believes the usual assumptions in deriving Newton's gravitational force law from general relativity are insufficient, that gases and stars in galaxies are not immune from Hubble flow. As a consequence a universal constant $a_{0}$ is introduced as a characteristic acceleration in the cosmos.  

Using spherical coordinates Carmeli \cite{Carmeli1998} successfully provided a theoretical description of the Tully-Fisher law for spiral galaxies. Following Carmeli's lead, Hartnett \cite{Hartnett2006} showed that the same line of reasoning leads to plausible galaxy rotation curves in spiral galaxies using cylindrical coordinates. 

In this paper I take the analysis further, using spherical coordinates and model the gravitational potential and the resulting forces determining how test particles move in spheroidal and elliptical galaxies with an appropriate density distribution that reflects the observed luminous matter distribution. Similar to what was found in spiral galaxies \cite{Hartnett2006}, two acceleration regimes are apparent here also. In one, normal Newtonian gravitation applies. In that regime the effect of the Hubble expansion is not observed or is extremely weak. It is as if particle accelerations are so great that they slip across the expanding space. In the other, new physics is needed. There the Carmelian metric provides it. In this regime the accelerations of particles are so weak that their motions are dominated by the Hubble expansion and as a result particles move under the combined effect of both the Newtonian force and a post-Newtonian contribution. 

\section{\label{sec:Gpotential}Gravitational potential}

In the weak gravitational limit, where Newtonian gravitation applies, it is sufficient to assume the Carmeli metric with non-zero elements $g_{00} = 1 + 2 \Phi /c^{2}$, $g_{44} = 1 + 2 \Psi /\tau^{2}$, $g_{kk} = -1$, ($k = 1, 2, 3$) in the lowest approximations in both $1/c$ and $1/\tau$. Here a new constant, called the Hubble-Carmeli constant, is introduced $\tau \approx 1/H_{0}$. See Ref.~\cite{Carmeli2002} for details. The potential functions $\Phi$ and $\Psi$ are determined by Einstein's field equations and from their respective Poisson equations,
\begin{eqnarray} 
\nabla^{2} \Phi = 4 \pi G \rho \label{eqn:phi},
\\
\nabla^{2} \Psi = \frac{4 \pi G \rho}{a_{0}^{2}} \label{eqn:psi},
\end{eqnarray}
\\
where  $\rho$ is the mass density, $a_{0}$ a universal characteristic acceleration $a_{0} =c/\tau$ and $c$ is the speed of light in vacuo. 

A comparison of $\Phi$ and $\Psi$ in (\ref{eqn:phi}) and (\ref{eqn:psi}) leads to $\Psi = \Phi /a_{0}^{2}$ within an arbitrary additive constant. Since both potentials are defined with respect to the same co-ordinate system, in reality, we only need deal with one potential function, the gravitational potential, $\Phi$.

The density function that best describes the luminous distribution of matter in spherical and elliptical galaxies is
\begin{equation} \label{eqn:rdensity}
\rho(r) = \frac{M}{4 \pi} \frac{r_c}{r^2(r+r_c)^{2}},
\end{equation}
where $r_c$ is the core radius which contains half of the luminous matter. The parameter $M$ is the total mass of the galaxy as measured at $r = \infty$. Therefore the mass at radius $r$ is
\begin{equation} \label{eqn:mass}
M(r) = M \frac{r}{r+r_c}.
\end{equation}
Spherical symmetry has been assumed to simplify the problem. To consider fully oblate spheroids would introduce some asymmetry to the problem but in principle the solution found here should broadly apply to any spherical matter distribution and would also apply, to first order, to oblate spheroids.

In spherical coordinates ($r,\theta, \phi$) the potential $\Phi$ that satisfies (\ref{eqn:phi}) can be found by integrating over (\ref{eqn:phi}) using (\ref{eqn:rdensity}). Only the radial dependence remains.
\begin{equation} \label{eqn:potential}
\Phi (r) = \frac{G M}{r_c} \left\{log \left(\frac{1+r_c/\Delta}{1+r_c/r}\right) -\frac{r_c/\Delta}{1+r_c/\Delta}\right\}
\end{equation}
where $\Delta$ is the radial extent of the matter distribution. This may be approximated where $\Delta \gg r_c$, which is the usual case. Hence we get,
\begin{equation} \label{eqn:approxpotential}
\Phi (r) = -\frac{G M}{r_c} log \left(1+\frac{r_c}{r}\right).
\end{equation}

\section{\label{sec:Eqnmotion}Equations of Motion}

The 5D line element for any two points in the CGR theory is $ds^{2} = g_{00} c^{2} dt^{2}+ g_{kk} (dx^{k})^2+ g_{44} \tau^{2} dv^{2}$, where $k = 1, 2, 3$. The relative separation in 3 spatial coordinates $r^{2} = (x^{1})^{2} + (x^{2})^{2} + (x^{3})^{2}$ and the relative velocity between points connected by $ds$ is $v$. The Hubble-Carmeli constant,  $\tau$, is a universal constant for all observers. The equations (B.62a) and (B.63a) from Ref.~\cite{Carmeli2002} are used in the sequel to derive the appropriate equations of motion to lowest approximation in $1/c$.

\subsection{\label{sec:Nequation}Newtonian}

It follows from Carmeli's equation (B62a)  using (\ref{eqn:rdensity}) and (\ref{eqn:potential}), and  the usual form of the circular motion equation 
\begin{equation} \label{eqn:Newtonian}
\frac{v^{2}}{r} = \frac{d \Phi} {dr},
\end{equation}
that
\begin{equation} \label{eqn:Newtonian2}
v^{2} = \frac{G M} {r}\frac{1}{1+r_c/r},
\end{equation}
where $G$ denotes the gravitational constant. 

\begin{figure}
\includegraphics[width = 5 in]{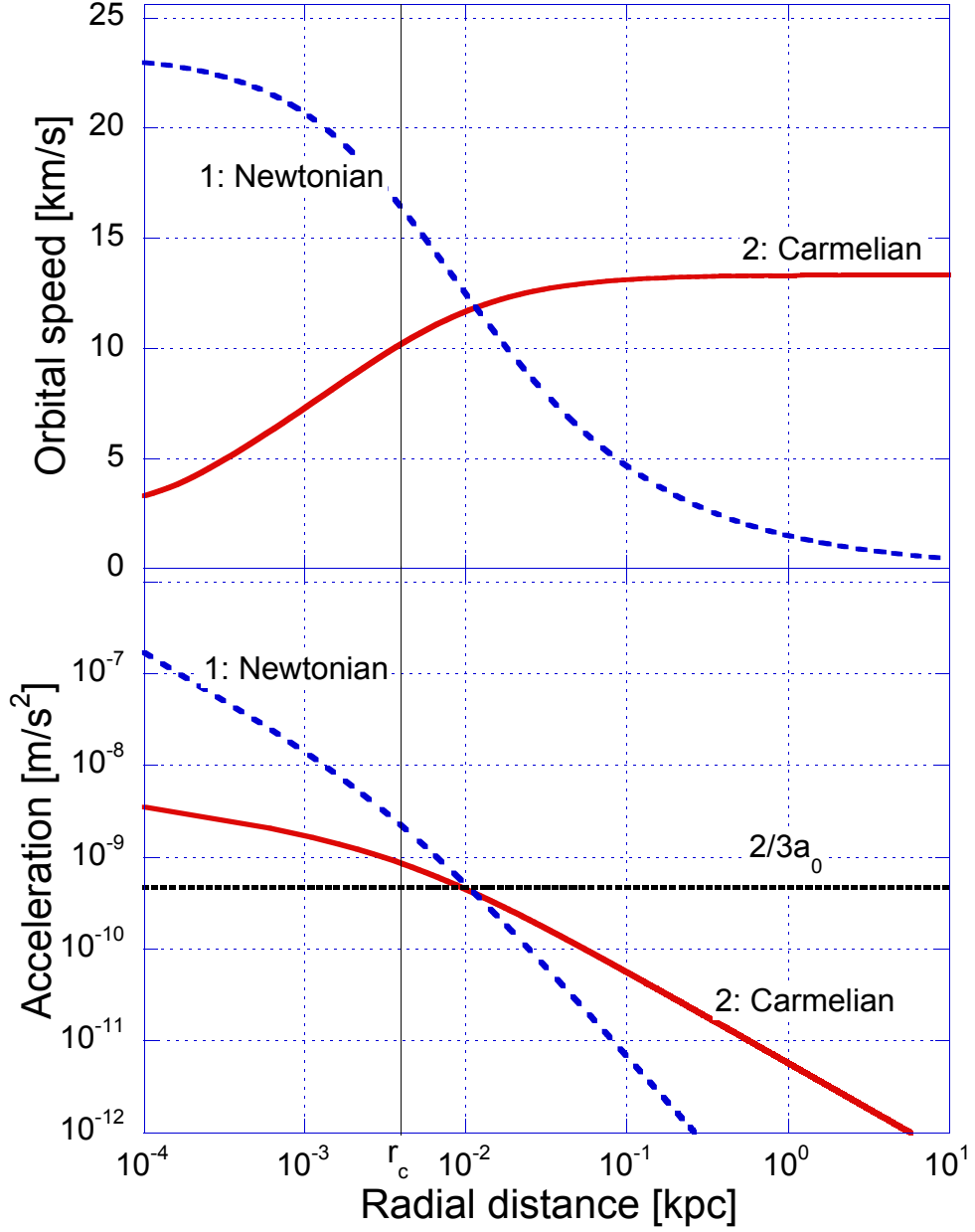}
\caption{\label{fig:fig1} (a) Orbital velocity of a test particle in a dwarf spheroidal galaxy calculated from (\ref{eqn:Newtonian2}) and (\ref{eqn:Carmelieqn2}) and where $M = 10^6 M_{\odot}$  and $r_c = 0.004\; kpc$, indicated by the black vertical line; (b) Corresponding acceleration of a test particle for both the Newtonian and Carmelian regimes. Note they coincide at the critical acceleration $2/3a_0$.}
\end{figure}

Equation (\ref{eqn:Newtonian2}) is the usual Newtonian result for the speed of circular motion in a spherical gravitational potential. This equation has been plotted in curve 1 of fig. \ref{fig:fig1}(a) as a function of radial position from the center of a galaxy in kiloparsecs ($kpc$) where $kpc \approx 3.08 \times 10^{19}$ m. Figure \ref{fig:fig1}(b) shows the corresponding accelerations. Curve 1 is the Newtonian acceleration. Through out this paper $M$ is expressed in solar mass units $M_{\odot} \approx 2 \times 10^{30}$ kg.

\subsection{\label{sec:Cequation}Carmelian}

Using $\Psi = \Phi /a_{0}^{2}$ in Carmeli's (B63a) results in a new equation
\begin{equation} \label{eqn:Carmelieqn}
v = a_{0}\int^{r}_{0}\frac{dr}{\sqrt{-\Phi}},
\end{equation}
which has been integrated and solved for $v$ as a function of $r$. Using the potential $\Phi$, determined from (\ref{eqn:rdensity}) and (\ref{eqn:potential}), in (\ref{eqn:Carmelieqn}), results in 
\begin{equation} \label{eqn:Carmelieqn2}
v = \frac{2}{3}a_{0} \frac{r^{3/2}}{\sqrt{GM}},
\end{equation}
which describes the expansion of space within a galaxy. 

In Refs~\cite{Carmeli1998} and \cite{Carmeli2000}, using spherical co-ordinates, it was found that in the limit of large $r$ and where all the matter was interior to the position of a test particle, such a particle is also subject to an additional circular motion described by (\ref{eqn:Carmelieqn2}). Apparently this is the result of the expansion of space itself within the galaxy but in an azimuthal direction to the usual center of coordinates of the galaxy. In this paper also the same result (\ref{eqn:Carmelieqn2}) was obtained.

Carmeli \cite{Carmeli1998, Carmeli2000} determined a Tully-Fisher type relation using the Newtonian circular velocity equation expressed in spherical coordinates,
\begin{equation} \label{eqn:Newtonianspherical}
v^{2} = \frac{G M }{r},
\end{equation}
where it is assumed that test particles orbit at radius $r$ outside of a fixed mass $M$. Then by eliminating $r$ between (\ref{eqn:Newtonianspherical}) and (\ref{eqn:Carmelieqn2}) we get the result. This is achieved by taking the $3/2$ power of (\ref{eqn:Newtonianspherical}) and multiplying it with (\ref{eqn:Carmelieqn2}) yielding
\begin{equation} \label{eqn:TF}
v^{4} = G M \frac{2}{3}a_{0}.
\end{equation}

So by applying the same approach with (\ref{eqn:Newtonian2}), raising it to the $3/2$ power and multiplying it with (\ref{eqn:Carmelieqn2}), we can derive an equation describing the circular motion of test particles in spherical galaxies. The result is 
\begin{equation} \label{eqn:rotcurve}
v^{4} = G M \frac{2}{3}a_{0} \left(\frac{r}{r+r_c} \right)^{3/2}.
\end{equation}
But in elliptical and spherical galaxies we generally don't observe any group circular motion. Therefore the trajectories of the individual orbits, though circular around the center of mass, are randomly oriented. Nevertheless we can calculate the orbital speed of a test particle from (\ref{eqn:rotcurve}). Curve 2 of fig. \ref{fig:fig1}(a) shows one for the same parameters as is assumed for the Newtonian curve, as a function of radial position from the center of a galaxy. However for an individual particle if $r \gg r_c$ equation (\ref{eqn:rotcurve}) recovers the form of the Tully-Fisher relation (\ref{eqn:TF}).

The corresponding acceleration may also be calculated from $v^2/r$ using (\ref{eqn:rotcurve}) and is shown in curve 2 of fig. \ref{fig:fig1}(b) for the same galaxy parameters. It is also compared with the critical acceleration $2/3a_0$.

\section{\label{sec:Dispersion}Radial velocity dispersion}

Assuming an spherically isotropic dynamic cloud of gases and stars, where the radial velocity dispersion is equal to the $\theta$ or $\phi$ dispersion (i.e. $\sigma^2_r= \sigma^2_{\theta, \phi}$) the hydrostatic Jeans equation becomes
\begin{equation} \label{eqn:Jeanseqn}
\frac{d}{dr}(\rho(r)\sigma^2_r) = -\rho(r) \frac{d\Phi}{dr}.
\end{equation}  

\subsection{Newtonian}

\begin{figure}
\includegraphics[width = 5 in]{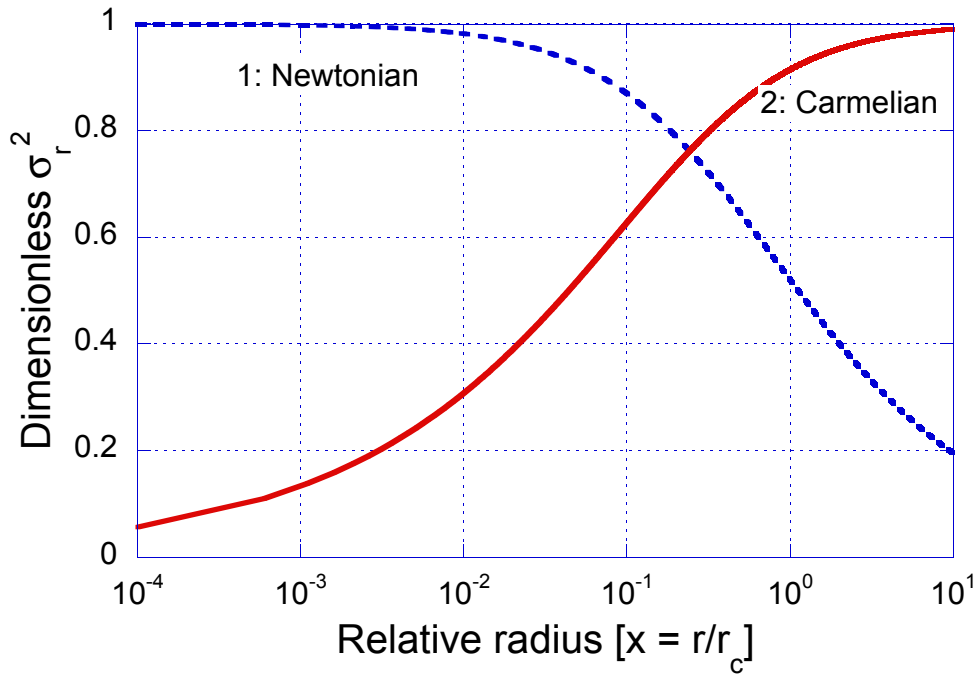}
\caption{\label{fig:fig2} Dimensionless velocity dispersion squared. Curve 1 is $\Xi(x)$ from (\ref{eqn:Xi}) and curve 2 is $\Gamma(x)$ from (\ref{eqn:Gamma}) as functions of the relative radius $x = r/r_c$. }
\end{figure}

Using (\ref{eqn:Newtonian}), (\ref{eqn:Newtonian2})  and (\ref{eqn:rdensity}) we solve the differential equation (\ref{eqn:Jeanseqn}) for the radial velocity dispersion 
\begin{equation} \label{eqn:Nsigmar}
\sigma^2_r = \frac{1}{2} \frac{GM}{r_c} \Xi(x),
\end{equation}
where 
\begin{equation} \label{eqn:Xi}
\Xi(x) = (1+2x)(1-6x(1+x))+12x^2(1+x)^2log\left(\frac{x}{1+x}\right)
\end{equation}
and $x = r/r_c$. The dimensionless function $\Xi(r/r_c)$ is shown as curve 1 in fig. \ref{fig:fig2}.

For $r \gg r_c$ the function $\Xi(x) \approx 2/5x$ which means (\ref{eqn:Nsigmar}) becomes
\begin{equation} \label{eqn:Nsigmarapprox}
\sigma^2_r = \frac{GM}{5r}.
\end{equation}
This is the same result determined from the virial theorem.

So in the limit where $r \rightarrow \infty$, the function $\Xi \rightarrow 0$ and where $r \rightarrow 0$, the function $\Xi \rightarrow 1$. From Newtonian theory we expect that the radial velocity dispersion in spheroidal and elliptical galaxies to tend to zero where the gravitational acceleration is very weak. And where the gravitational acceleration is strongest ($r = 0$) the radial dispersion is maximum and equal to the rotation velocity of a particle at twice the core radius.

\subsection{Carmelian}

Using (\ref{eqn:rotcurve}) we can construct an equivalent gravitational potential which includes the effects of the Hubble flow on test particles in the galaxy. By taking $v^2/r$ and using this in (\ref{eqn:Jeanseqn}) instead of the gradient of the Newtonian potential we get
\begin{equation} \label{eqn:Csigmar}
\sigma^2_r = \frac{1}{4} \left(G M \frac{2}{3}a_{0}\right)^{1/2}  \Gamma(x),
\end{equation}
where
\begin{eqnarray}
\Gamma(x) = \frac{16}{35}x^{3/4}(1+x)^{1/4}\{7-4x (21-32(x(1+x)^3)^{1/4}+... \\ \nonumber
...+ 8x(7+4x-4(x(1+x)^3)^{1/4}) ) \}, \label{eqn:Gamma}
\end{eqnarray}
and $x = r/r_c$. See curve 2 in fig. \ref{fig:fig2} for $\Gamma(r/r_c)$.

In the limit where $r \rightarrow \infty$, the function $\Gamma \rightarrow 1$ and where $r \rightarrow 0$, the function $\Gamma \rightarrow 0$. From Carmelian theory then we expect that the radial dispersion in spherical and elliptical galaxies to tend to zero where the gravitational acceleration is strongest ($r = 0$) and to tend to a constant where the gravitational acceleration is very weak. It also follows from (\ref{eqn:Csigmar}) that for $r \gg r_c$
\begin{equation} \label{eqn:Csigmar4}
\sigma^4_r = \frac{1}{16} \left(G M \frac{2}{3}a_{0}\right).
\end{equation}

\subsection{Discussion}

Similarly to what was seen in the case of rotation curves \cite{Hartnett2006}, two regimes must be considered. When accelerations are greater than the critical acceleration $2/3a_0$ we have the Newtonian regime and we expect the velocity dispersion to follow (\ref{eqn:Nsigmar}) but when accelerations are less than $2/3a_0$ we have the Carmelian regime  and we expect the velocity dispersion to follow (\ref{eqn:Csigmar}). Notice that even though the velocity dispersion from the Newtonian calculation (\ref{eqn:Nsigmar}) scales with core radius ($r_c$), the velocity dispersion from the Carmelian calculation (\ref{eqn:Csigmar}) is independent of core radius.

\begin{figure}
\includegraphics[width = 5 in]{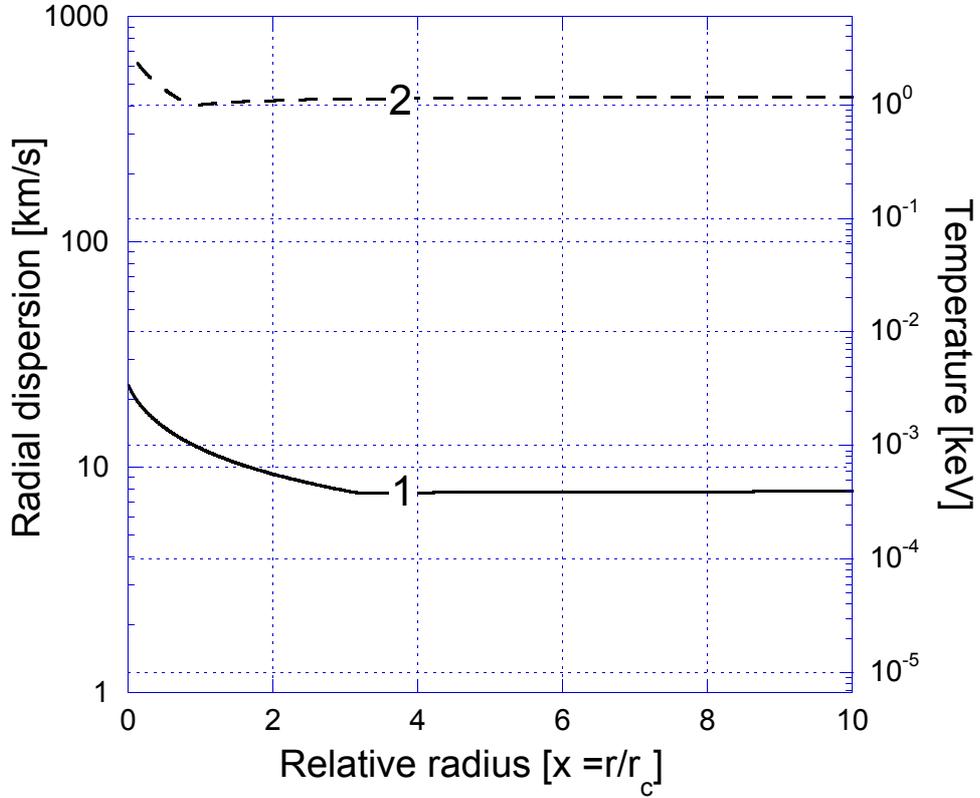}
\caption{\label{fig:fig3} Left axis: Radial velocity dispersion as a function of $x = r/r_c$. Right axis: Temperature of gas in units of $keV$. Curve 1 is $\sigma_r(x)$ for a dwarf spheroidal galaxy with $M = 10^6 \;M_{\odot}$ and curve 2 is $\sigma_r(x)$ for a massive elliptical galaxy with $M = 10^{13} \;M_{\odot}$. }
\end{figure}

The Newtonian regime applies where the Newtonian acceleration $a_N > 2/3a_0$ and the Carmelian regime applies where the Carmelian acceleration $a_C < 2/3a_0$. In fig. \ref{fig:fig3} radial velocity dispersion curves are shown for a dwarf spheroidal (curve 1) with $M = 10^6 \;M_{\odot}$  and a massive elliptical (curve 2) with $M = 10^{13} \;M_{\odot}$. Only the regions of the curves which represent valid velocity dispersion are retained. The results indicate that, as a function of radius from the center of the galaxy, the velocity dispersion should fall from a central high value to a constant value as $r$ becomes much greater than $r_c$. This is commonly observed in these type of galaxies.

\subsubsection{Dwarf spheroidal galaxies}

It follows from (\ref{eqn:Csigmar4}), that we can write an equation for the radial velocity dispersion that we expect to observe in a typical dwarf galaxy, of mass $M$ (in units of $M_{\odot}$) as
\begin{equation} \label{eqn:CsigmaDwarf}
\sigma_r = 4.40  \left(\frac{M}{10^5 M_{\odot}}\right)^{1/4} \, km/s.
\end{equation}
assuming the observations are obtained from the sources far from the core radius. That is, sources that are in the Carmelian acceleration regime. Here a value of $\tau = 4.276 \times 10^{17}$ s has been assumed \cite{Oliveira2006}, which is equivalent to $\tau^{-1} = 72.17$ \kmsmpc, and hence $a_0 = 4.674 \times 10^{-10}$ m/s .

By calculating the mass required for typically observed values of $\sigma_r$, in dwarf spheroidal galaxies, according to the both Newtonian (\ref{eqn:Nsigmar}) and Carmelian equations (\ref{eqn:Csigmar}) developed here a comparison can be made. The galaxy mass ($M$) calculated for different radial velocity dispersions  $\sigma_r$ at $x = r/r_c = 10$ (for a typical core radius of $r_c = 0.004 \;kpc$) are listed in Table I.

\begin{table}[ph]
\begin{center}
Table I: Masses of spheroidal dwarf galaxies from \\velocity dispersion $\sigma_r$ at $x = r/r_c = 10$
\vspace{6pt}
\begin{tabular}{c|ccc} \hline
\hline
 $\sigma_r$	& Carmelian												& Newtonian													&  \\
 $[km/s]$		& $M [\times 10^{5} M_{\odot}]$		& $M [\times 10^{5} M_{\odot}]$ 		& Ratio\\
\hline
4.40 				& 1.00 						& 9.38 				& 9.38 \\
6.58				& 5.00  					& 20.9 				& 4.20 \\
7.82				& 10.0 						& 29.7 				& 2.96 \\
11.7				& 50.0 						& 66.2  			& 1.33 \\
\hline
\hline
\end{tabular}
\vskip 2mm 
\end{center}
\end{table}

\subsubsection{Bright central elliptical galaxies}

Similarly from (\ref{eqn:Csigmar4}) we can write an equation for for the radial velocity dispersion that we expect to observe in a typical massive elliptical galaxy, of mass $M$ (in units of $M_{\odot}$) as
\begin{equation} \label{eqn:CsigmaEllip}
\sigma_r = 247.4  \left(\frac{M}{10^{12} M_{\odot}}\right)^{1/4} \, km/s.
\end{equation}
also assuming the observations are obtained from the sources that are far from the core radius, in the Carmelian acceleration regime.

Again by calculating the mass required for typically observed values of $\sigma_r$, in bright ellipticals using both models a comparison is made results listed in Table II. In this case a typical core radius of $r_c = 40 \;kpc$ was assumed. 

\begin{table}[ph]
\begin{center}
Table II: Masses of bright elliptical galaxies from \\ velocity dispersion $\sigma_r$ at $x = r/r_c = 10$
\vspace{6pt}
\begin{tabular}{c|ccc} \hline
\hline
 $\sigma_r$	& Carmelian													& Newtonian 												&  \\
 $[km/s]$		& $M [\times 10^{12} M_{\odot}]$		& $M [\times 10^{12} M_{\odot}]$ 		& Ratio\\
\hline
139.1					& 0.1 						& 9.38  			& 93.8 \\
247.4 				& 1.0 						& 29.7	 			& 29.7 \\
369.9					& 5.0  						& 66.3 				& 13.3 \\
439.9					& 10.0 						& 93.8 				& 9.38 \\
\hline
\hline
\end{tabular}
\vskip 2mm
\end{center}
\end{table}

\subsubsection{Temperature}

By considering the random motion of gas particles we can write an expression for the temperature of a cloud of gas in the galaxies that is heated by the gravitational potential the gas feels whether it be in the Newtonian or Carmelian regime. Using $\mu = 0.609$ as the mean atomic weight of the gas we can equate
\begin{equation} \label{eqn:temp}
\mu m_p \sigma_r^2 = k T,
\end{equation}
where $k$ is Boltzmann's constant, $m_p$ is the proton mass and $T$ is the temperature of the gas. Using (\ref{eqn:temp}) the temperature of the gas in a galaxy heated in this manner can be determined from the radial velocity dispersion. This is shown on the right axis of fig. \ref{fig:fig3}. The result indicates that the temperature of the gas across spherical galaxies is expected to be approximately isothermal. Also, due to the thermal heating of intra-galactic gas, massive elliptical and spheroidal galaxies emit radiation in the X-ray part of the spectrum, while dwarf spheroidals emit in the near infrared. This is consistent with observations.

\section{\label{sec:Conclusion}Conclusion}

This theory suggests that it is the Carmelian regime that is applicable at low accelerations where $r \gg r_c$ and hence the masses of galaxies are overestimated from the observed dynamics. In Table I and II typical dispersion velocities are used but the core radius is fixed at $r_c = 0.004 \; kpc$ and $r_c = 40 \; kpc$, respectively. So depending on the exact details for a galaxy one may get 10 to 100 times more mass from a Newtonian calculation than from a Carmelian calculation. This brings the masses more in line with estimates from the luminous material. Therefore the need to invoke halo dark matter is avoided. It is also true that $\sigma_r$ is observed to fall from a central higher value near the center of the spheroidal galaxy and become constant as a function of radius from the center. This is indicated by a Newtonian regime in the center which becomes Carmelian as a function of radius. In massive ellipticals radial velocity dispersion is often observed to be  approximately constant as a function of radius. This is indicated by observations where most of the galaxy is dominated by the Carmelian regime.

\section{Acknowledgment}
This work was supported by the Australian Research Council.


\begin{thebibliography}{99}

\bibitem{Carmeli1982} Carmeli, M. (1982). \textit{Classical Fields, General Relativity and Gauge Fields}, New York, Wiley
\bibitem{Carmeli1998} Carmeli, M. (1998). Is galaxy dark matter a property of spacetime?, Int. J. Theor. Phys.  \textbf{37} (10): 2621-2625
\bibitem{Carmeli2000} Carmeli, M. (2000). Derivation of the Tully-Fisher Law: Doubts About the Necessity and Existence of Halo Dark Matter, Int. J. Theor. Phys. \textbf{39} (5): 1397-1404
\bibitem{Carmeli2002} Carmeli, M. (2002). \textit{Cosmological Special Relativity}. Singapore, World Scientific
\bibitem{Hartnett2006} Hartnett, J.G. ``Spiral galaxy rotation curves determined from Carmelian general relativity'' \textit{Int. J. Theor. Phys.} \textbf{45}(11):2147--2165 (2006) 
\bibitem{Oliveira2006} Oliveira, F.J., J.G. Hartnett,  ``Carmeli's cosmology fits data for an accelerating and decelerating universe without dark matter or dark energy'' \textit{Found. Phys. Lett.} \textbf{19}(6):519--535 (2006)  

\end{thebibliography}
\end{document}